\documentclass[%
 aip,
cp,  % Conference Proceedings
 amsmath,amssymb,%nobibnotes,
reprint,%
]{revtex4-2}

\usepackage{graphicx}% Include figure files
\usepackage{dcolumn}% Align table columns on decimal point
\usepackage{bm}% bold math
\usepackage[utf8]{inputenc}
\usepackage[T1]{fontenc}
\usepackage[english]{babel}

\makeatletter
\let\ORIbbl@fixname\bbl@fixname
\def\bbl@fixname#1{%
  \@ifundefined{languagealias@\expandafter\string#1}
    {\ORIbbl@fixname#1}
    {\edef\languagename{\@nameuse{languagealias@#1}}}%
}
\newcommand{\definelanguagealias}[2]{%
  \@namedef{languagealias@#1}{#2}%
}
\makeatother

\definelanguagealias{en}{english}

\usepackage{mathptmx}

\newcommand{\vecb}[3]{
\bm{\vec{#1}_{#2}^{#3}}
}

\newcommand{\uvecb}[3]{
\bm{\hat{#1}_{#2}^{#3}}
}

\begin{document}

\title{MIDSX: A Monte Carlo Interaction and Dosimetry Simulation of X-rays}% Force line breaks with \\

\author{John Meneghini} % Write as First name Surname
 \email[Corresponding author: ]{john.meneghini@stvincent.edu}
\affiliation{
 Department of Physics, Saint Vincent College, Latrobe, PA 15650
}

\date{11/28/23} % It is always \today, today, but any date may be explicitly specified
              % Not printed for conference proceedings

\begin{abstract}
Computed Tomography (CT) imaging, while essential for diagnostics, exposes patients to ionizing radiation. To accurately quantify radiation dosage, this study introduces MIDSX, a specialized open-source Monte Carlo (MC) photon transport code system for medical imaging. Unlike general purpose particle transport MC systems, MIDSX is tailored for x-ray transport, offering more streamlined implementation. Results from validation simulations show MIDSX's results for specific cases agree to within 0.39\% of the mean of established reference data. However, discrepancies in body energy deposition measurements, reaching up to a 6.3\% mean percent error, indicate areas for further refinement in the system.
\end{abstract}

\maketitle

\section{INTRODUCTION}
\par Computed Tomography (CT) imaging is a critical diagnostic tool used by medical professionals to diagnose various illnesses and injuries. While the use of CT imaging is essential to provide immediate, life-saving results, ionizing radiation can damage cells and increase the risk of cancer. While the risk is small, it is cumulative, so physicians must track a patient's radiation exposure over time \cite{lauer2009elements}. Typically, exposure is measured using absorbed dose in the body and air kerma at the skin layer, which both have units of grays (joule/kg), making it directly related to the energy deposited by photons and their secondary particles. Often, these values are estimated based on the properties of the x-ray source and the specific procedure being performed. However, the most accurate estimation techniques use Monte Carlo (MC) methods to simulate the propagation of photons through a computational phantom \cite{essmedphys2012}. 
\par In the MC technique, the 3D space encompassing the phantom and the radiation source is represented as a computational domain. Within this domain, individual photon interactions are stochastically simulated, accounting for each interaction event as photons navigate the phantom. Such stochastic simulation offers unparalleled precision, capturing even the most subtle nuances of radiation behavior in biological media. Additionally, the MC method can simulate various medium types, densities, and configurations, making it incredibly versatile and adaptable to various imaging tests. Furthermore, advancements in computational power and algorithms have expedited MC simulation, rendering it more accessible and feasible for routine clinical applications \cite{fernandez_bosman_validation_2021}.
\par In this paper, the newly developed, open-source MC photon transport code system MIDSX is presented and validated. While many existing MC transport code systems perform reliably in dosimetry applications \cite{fernandez_bosman_validation_2021, geant4valid2004}, many of these systems are tailored for general particle transport. The developmental focus of MIDSX on x-ray transport reduces the complexity of implementation and allows users to easily design and run simulations specifically relating to x-ray transport in the medical imaging energy range. The subsequent sections will delve into the theory of MIDSX, compare results from MIDSX to accepted benchmarks from established simulation systems, and outline future work.

\section{TRANSPORT THEORY}
\par To represent the computational domain discretely, space is broken up into a grid of voxels (volume pixel), with each voxel being assigned a particular material depending on the geometries and compounds/elements in the domain. Within our discrete space, given a photon position $\vecb{r}{}{}$, the corresponding voxel in which the photon resides can be calculated. Therefore, all possible $\vecb{r}{}{}$'s can be assigned a particular material $M$ in the domain.

\par A photon's position after taking the $n$-th step in the domain, $\vecb{r}{n}{}$, is represented by the parametric ray equation:
\begin{equation} \label{eq:1}
    \vecb{r}{n}{} = \vecb{r}{n-1}{} + \uvecb{d}{}{} t_n,
\end{equation}
where $\vecb{r}{n-1}{}$ is the initial position before the $n$-th step, $\uvecb{d}{}{}$ is a unit vector in the direction of the step, and $t_n$ is the length of the $n$-th step.

\par To randomly sample $t_n$ in a homogeneous domain, we utilize the following probability density function (PDF) $p(t)$ of the distance traveled $t$ by a photon of energy $E$ through material $M$ before interacting:
\begin{equation} \label{eq:2}
    p(t) = n\sigma \exp{\left[-t(n\sigma)\right]},
\end{equation}
where $n$ is the number density of $M$ and $\sigma = \sigma(E, M)$ is the microscopic cross-section of $M$ at $E$.
\par Using the inversion method for sampling a PDF on Eq. \ref{eq:2}, random values of the free path $t$ can be generated by $
    t = -\frac{1}{n\sigma} \ln \gamma, 
    $
    %inlining for space concerns
where $\gamma$ is a uniformly distributed random number in the interval $[0, 1)$. This value of $t$ is sampled for each step and is used as $t_n$ in Eq. \ref{eq:1} to determine the length of the $n$-th step \cite{vassiliev_monte_2017}.
\par If, after taking a step, the photon lands in a voxel with a different material (an inhomogeneous domain), then the corresponding free path for the new material must be calculated. To accomplish this, MIDSX employs $\delta$-tracking, which is significantly more computationally efficient for domains with similar cross-sections \cite{vassiliev_monte_2017}.

\par For x-rays, there are three possible photon interactions: photoelectric effect, coherent scattering, and incoherent scattering. For the photoelectric effect in MIDSX, the photon is terminated and all energy is deposited at the location of interaction. In general-purpose particle transport code systems, when a photoelectric interaction occurs, a photon of energy $E$ is absorbed by an electron in subshell $i$, causing the electron to leave the atom with energy $E_e = E - U_i$, where $U_i$ is the binding energy of the $i$th subshell. In addition, photons are emitted due to atomic relaxations. For photon energies in the medical imaging range (<120 keV), the energy of the released electrons does not allow for significant traversal through typical biological media. This limited traversal results in a localized dose distribution, in turn, validating the model used by MIDSX.

\par For coherent and incoherent scattering, the methodology of \cite{lund2018implementation} was adapted for use in MIDSX, neglecting Doppler energy broadening and the production of secondary particles.

\section{Results}
\par To validate the accuracy of MIDSX, simulations were performed and compared to reference data (PENELOPE, EGSrnc, Geant4, and MCNP) obtained by the American Association of Physicists in Medicine Task Group Report 195 (TG-195) \cite{sechopoulos_monte_2015}. The simulations performed from TG-195 were Case 1: "Half Value Layer," Case 2: "Radiography and Body Tomosynthesis," and Case 5: "CT with a Voxelized Solid." For Case 1, the primary air kerma was measured on a far away, circular region of interest (ROI) with a cone beam point source collimated such that all primary particles would be incident upon the ROI. The primary air kerma was measured with the domain filled only with air and then compared to the measured air kerma with an aluminum filter of thickness $t$ placed between the source and ROI. The ratios of the half value layer (HVL) and quarter value layer (QVL) primary air kerma to the primary background air kerma is represented by $R_1$ and $R_2$, respectively. By setting $t$ to correspond to the HVL and QVL for a particular spectrum, one can validate the material attenuation properties of an MC code system by comparing the simulated $R_1$ and $R_2$ to their theoretical values of 0.5 and 0.25, respectively. The simulation was performed for the monoenergetic energies of 30 keV and 100 keV, along with the polyenergetic spectrums of 30 kVp and 100 kVp, which were provided by TG-195. The term kVp (kilovolt peak) refers to the maximum voltage applied to an x-ray tube, which determines the highest energy of the x-rays produced. MIDSX's results for Case 1 agree to within 0.39\% of the mean results published by TG-195 (not shown).
\par
\begin{figure}[p]
    \centering
	\includegraphics[width=0.90\textwidth]{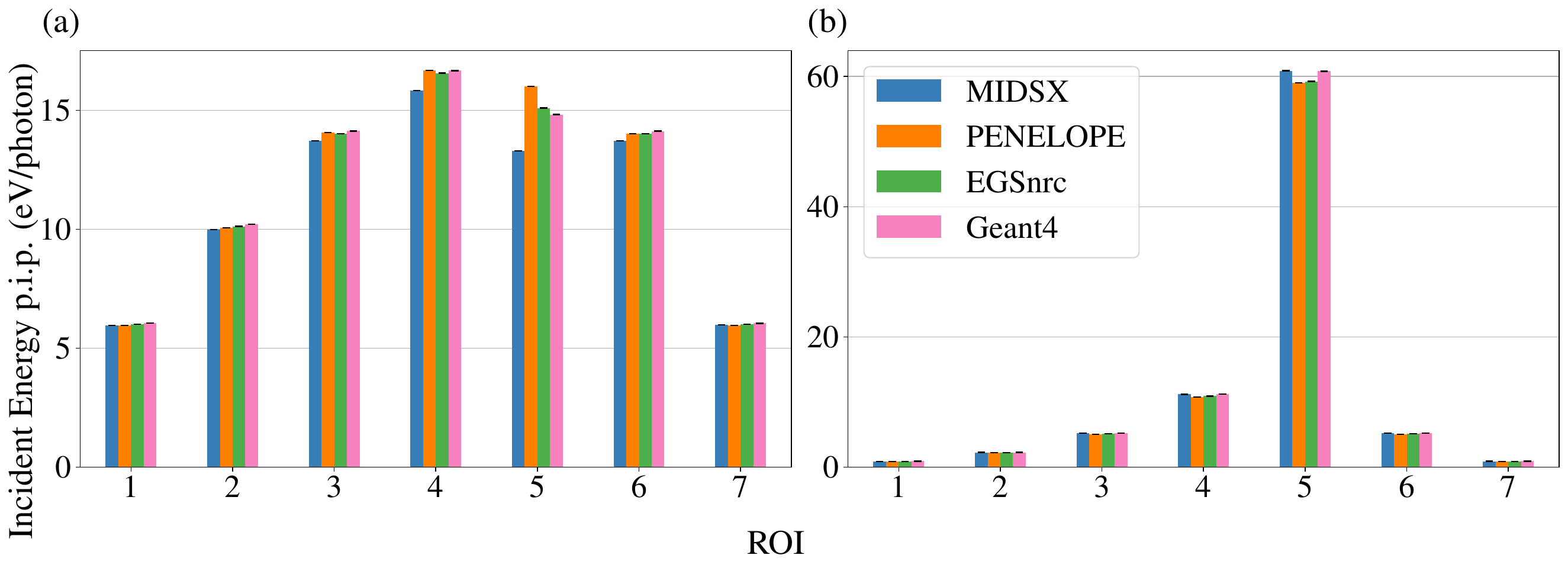}
	\caption{The energy per initial photon (p.i.p.) (eV/photon) of photons incident upon each region of interest (ROI) for the $0^\circ$, pencil beam, 56.4 keV simulation as described by Case 2. The incident energy was determined separately for photons that underwent (a) a single incoherent scatter and (b) a single coherent scatter.}
	\label{fig:ROIPGraph}
\end{figure}

\begin{figure}[p]
    \centering
	\includegraphics[width=0.90\textwidth]{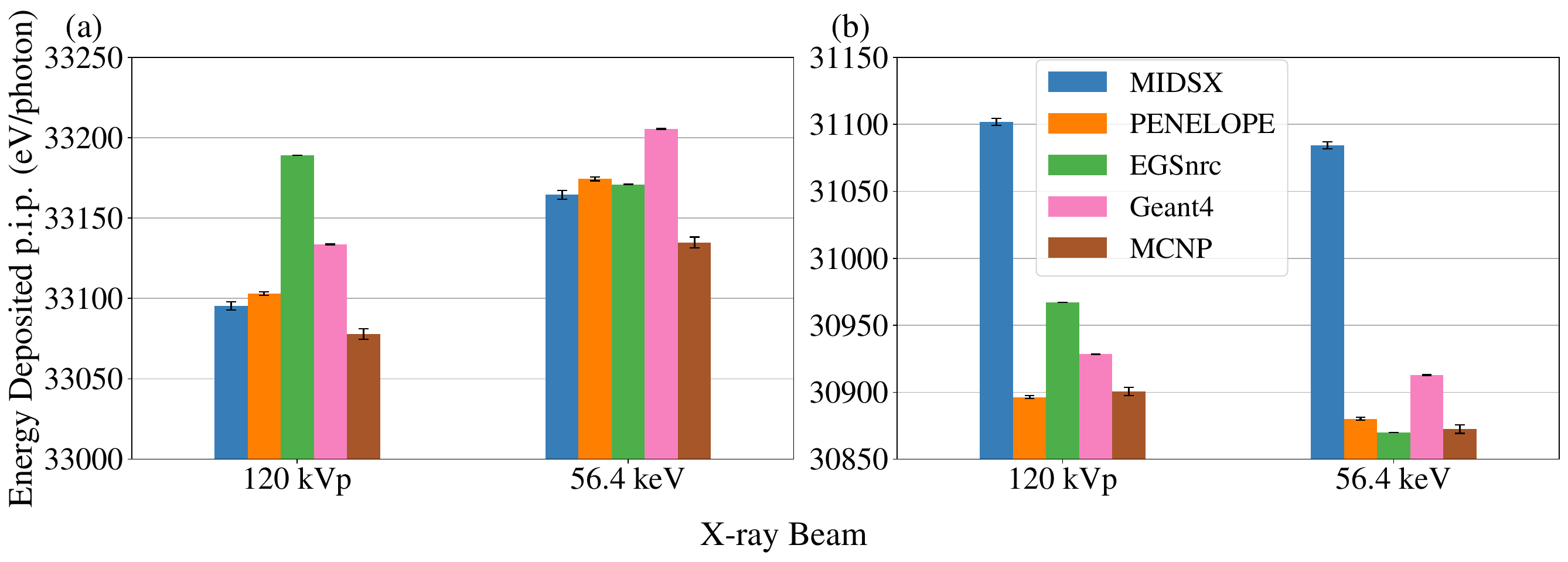}
	\caption{The energy deposited per initial photon (p.i.p.) (eV/photon) in the simulated tissue for the full-field simulation as described by Case 2. The simulation was performed at 56.4 keV and 120 kVp at both (a) $0^\circ$ and (b) $15^\circ$, with the 120 kVp spectrum provided by TG-195.}
 	\label{fig:BDGraph}
\end{figure}

 For Case 2, a full-field and pencil beam x-ray source were directed towards a cuboid tissue phantom at $0^\circ$ and $15^\circ$ from an axis drawn perpendicularly through one face from the center of the cuboid. Directly behind and inside the phantom, a grid of square ROIs and cube volumes of interests (VOI) were placed, respectively. The simulation was performed for the TG-195-provided polyenergetic spectrum of 120 kVp and its mean energy of 56.4 keV. For the $0^\circ$, full-field ROI measurements (not shown), a $<3.9$\% mean percent error (MPE) is seen for MIDSX's results to each ROI simulation. Furthermore, for the $0^\circ$ pencil-beam ROI measurements shown in Fig. \ref{fig:ROIPGraph}, a $<2.1$\% MPE is observed for each ROI simulation except for the case of a single incoherent scatter. In this particular case, MIDSX's results for ROI 4 and 5 are significantly lower, with the MPE reaching 13.1\% for ROI 5. The full-field VOI energy deposition measurements depicted in Fig. \ref{fig:BDGraph} show a minimal MPE of less than 0.1\% for the $0^\circ$ source. Conversely, for the MIDSX results at $15^\circ$, the MPE reaches an unexpectedly larger value of approximately 0.6\%.
\par
\begin{figure}[p]
    \centering
	\includegraphics[width=0.86\textwidth]{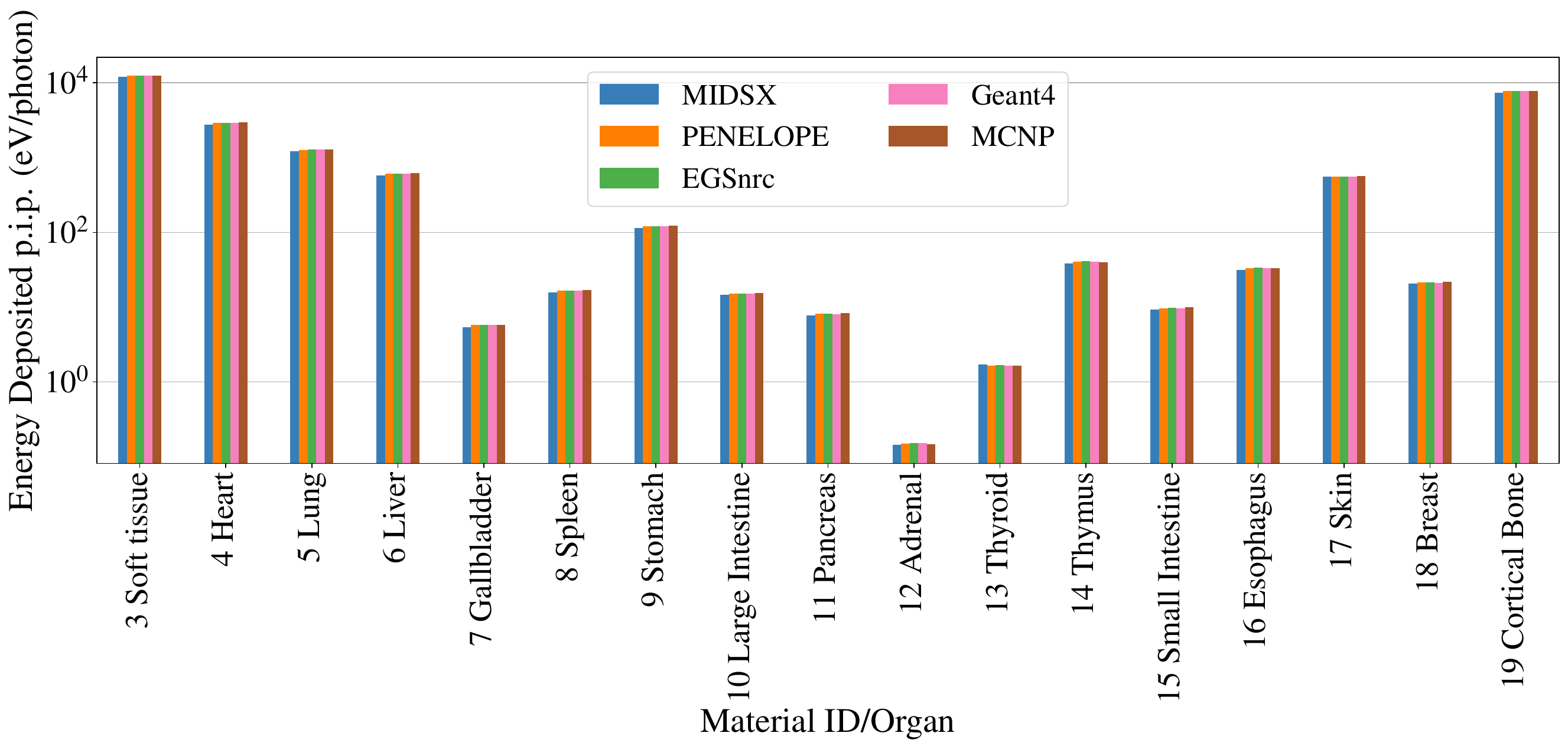}
	\caption{The energy deposited per initial photon (p.i.p.) (eV/photon) in the material IDs/organs composing a voxelized human phantom for the $0^\circ$, 120 kVp simulation as described by Case 5.}
	\label{fig:CTGraph}
\end{figure}

 For Case 5, a fan beam was collimated to the center of a voxelized human torso phantom provided by TG-195. To replicate a CT image, the simulation was repeated for several angles along a circle surrounding the phantom. The simulation was performed with Case 2's 120 kVp energy spectrum, and energy deposition was measured in the different materials/organs composing the phantom. Almost all of MIDSX's results for the $0^\circ$ source presented in Fig. \ref{fig:CTGraph} are systematically lower than the mean of the reference code systems, with MPE's ranging from 1.1\% to 6.3\%. This pattern is disrupted by the thyroid, which is larger than the mean by 2.9\%. 

\newpage
\section{Discussion}
\par Overall, MIDSX shows varied but reasonable agreement with the reference code systems of TG-195. For Case 1, excellent agreement is observed, in effect validating the total and max cross-section data used in $\delta$-tracking and the sampling of a discrete x-ray energy spectrum. This demonstrates that MIDSX is a reliable and accurate option for primary particle measurements.

\par For the ROIs of Case 2, agreement is seen almost universally, except for the single incoherent scatter deposition energy in ROI 4 and 5 for the pencil beam source shown in Fig. \ref{fig:ROIPGraph}. In particular, the results reached a max MPE of 13.1\% for ROI 5, indicating a potential error in the incoherent scattering energy/angular distribution sampling algorithm. This discrepancy is likely a result of an error in the rejection sampling algorithm employed by MIDSX. While this algorithm shows agreement for the full-field ROI 5, the geometry of the ROI, combined with the pencil beam, results in only narrow-angle scatters hitting the ROI. Since the scattering angle distribution of incoherent scattering at the medical imaging energy range becomes extremely steep at the scattering angle $\theta = 0^\circ$, there is likely numerical instability in the algorithm that needs to be analyzed. In addition, the $15^\circ$ full-field tissue energy deposition measurements were larger than the reference code systems', with an MPE of 0.6\%. With the MPE increasing significantly from the $0^\circ$ to $15^\circ$ measurements, this hints at a possible geometric error with the source and/or body. However, the source's position and angular distribution, along with the domain's and tissue's dimensions, were carefully verified, making the scene geometry unlikely to be the source of discrepancy.

\par For Case 5, almost all organ energy deposition results were lower than the reference code results, with the MPE reaching 6.3\%, except for the thyroid, with an MPE of 2.9\% larger. One common error reported by TG-195 that could result in the observed discrepancies is the incorrect orientation of the voxelized phantom in the computational domain. The orientation was verified by taking the root mean square percent error (RMSPE) of the MIDSX data with respect to the results of each simulated angle reported by TG-195. As expected, the RMSPE with respect to $0^\circ$ was the minimum, verifying that the phantom's orientation during the CT simulation was correct.

\par Despite verifying the orientation, the deviation of MIDSX's energy deposition results for both Cases 2 and 5 suggest that there may be other underlying errors in the MIDSX system that need further investigation. Potential factors could include the software's handling of scattering events, cross-section data initialization, and interpolation. Future work will study these aspects to pinpoint and rectify the source of the systematic errors observed in the MIDSX results.
\par The source code of MIDSX and examples using the system can be found in its GitHub repository \cite{MIDSX2023}.

\begin{acknowledgments}
    I would like to acknowledge my advisor, Fr. Michael Antonacci, for his invaluable guidance throughout my academic career. Immense gratitude to the A.J. and Sigismunda Palumbo Charitable Trust for their generous grant.
\end{acknowledgments}

% \nocite{*}
\bibliography{main}

\end{document}